# Defining a Simulation Strategy for Cancer Immunocompetence

Grazziela P. Figueredo' and Uwe Aickelin'

Intelligent Modelling and Analysis Research Group,
School of Computer Science,
The University of Nottingham, NG8 1BB, UK
gzf, uxa@cs.nott.ac.uk

**Abstract.**
Although there are various types of cancer treatments, none of these currently take into account the effect of ageing of the immune system and hence altered responses to cancer. Recent studies have shown that *in vitro* stimulation of T cells can help in the treatment of patients. There are many factors that have to be considered when simulating an organism's immunocompetence. Our particular interest lies in the study of loss of immunocompetence with age. We are trying to answer questions such as: Given a certain age of a patient, how fit is their immune system to fight cancer? Would an immune boost improve the effectiveness of a cancer treatment given the patient's immune phenotype and age? We believe that understanding the processes of immune system ageing and degradation through computer simulation may help in answering these questions. Specifically, we have decided to look at the change in numbers of naive T cells with age, as they play a important role in responses to cancer and anti-tumour vaccination. In this work we present an agent-based simulation model to understand the interactions which influence the naive T cell populations over time. Our agent model is based on existing mathematical system dynamic model, but in comparisons offers better scope for customisation and detailed analysis. We believe that the results obtained can in future help with the modelling of T cell populations inside tumours.

## 1 Introduction

According to estimations, over a million cases of colorectal cancer are diagnosed annually. This type of cancer is also the second most common cause of cancer-related deaths. Despite many different types of cancer treatments, survival rates are low and remain between 50% and 10%, depending on the cancer stage. However, none of the actual treatments take into account the patient's immune fitness or the immune system's capability to respond to cancer.

Recent research [1, 2] shows success in laboratory stimulation of T cells that kill tumour cells. However, in real patients there are additional factors to be considered before the stimulation. It is necessary to understand the patient's ability to respond to the treatment. Clinical trials are important, but not necessarily the end point. Therefore further investigation is necessary to determine all the factors to be observed before anti-tumour vaccination. We believe that one of the factors that should be considered is immunocompetence with age. Therefore, we want to investigate if we can use computational simulation models to verify if the immune responses to cancer changes in different life stages.

With age there is a decay of the immune system's performance resulting in degenerative diseases and deregulated and ineffective responses. The ageing of the immune system is called immunosenescence. Following an extensive study of the immunosenescence literature, we point out the four most influential theories in section 2. All four theories seem to be good candidates for developing immunosenescence simulation models. The four theories are: lack of naive T cells, immunological space filling with memory cells, innate system up-regulation and accumulation of T-regulatory cells.

The objective of our work is to find out if there is the possibility to predict what actually is going to happen if a patient with a certain age is anti-tumour vaccinated. To build such a computer simulation, one of the above theories has to be chosen first, as they are not all mutually compatible. We will discuss later in more detail why we believe that the first theory is the most appropriate one to be used in cancer simulation models: T cell populations are a major contributor to the immune system's functionality. They also play a very important role in responses to cancer and anti-tumour vaccination.

In this paper, we present results of our simulation model showing the interactions which influence the naive T cell populations over time. The model is based on the mathematical equations defined in [3]. In their work, Murray et al. [3] propose a model with a set of equations to fit some observed data and they try to estimate the likely contribution of each of the naive T cell repertoire maintenance methods. Therefore, we believe that the results of this model could be used as an input to a bigger model involving T cells inside tumours.

The work is organized as follows. Section 2 presents the immunological concepts related to immune system ageing. Next, in Section 3, there is a summary of the main theories that could be used for computational modelling. In Section 4 we present the model studied and its results. Finally, we draw conclusions and present future steps of this research in Section 5.

## 2 Background

According to Bulati et al [4], ageing is a complex process that negatively impacts on the development of the immune system and its ability to function. Progressive changes of the innate and adaptive immune systems have a major impact on the capacity of an individual to produce effective immune responses.

The decrease of immunocompetence in the elderly can be envisaged as the result of the continuous challenge of the unavoidable exposure to a variety of potential antigens, e.g. viruses, bacteria, food and self-antigens [5]. Antigens are the cause of persistent life-long antigenic stress, responsible for the filling of the immunological space by an accumulation of effector T cells and immunological memory [5].

With age, there is also a significant reduction of naive T cells numbers caused by the involution of the thymus. This situation eventually leaves the body more susceptible to infectious and non-infections diseases [3]. Also, there is evidence that clonotypical immunity deteriorates, while ancestral innate or natural immunity is conserved or even up-regulated [5, 6].

## 2.1 Some Factors Related to Immunosenescence

According to Franceschi [5], some factors that characterise immunosenescence are the accumulation of memory T cells, the decrease and exhaustion of naive T cells and a marked reduction of the T cell repertoire. Bulati [4], on the other hand, believes that both innate and adaptive immunity are usually involved in the pathogenesis of chronic age-related diseases like arthritis, atherosclerosis, osteoporosis, diabetes and so on. However, the innate immune system appears to be the prevalent mechanism driving tissue damage associated with different age-related diseases [4]. Thus, ageing is accompanied by an age-dependent up-regulation of the inflammatory response, due to the chronic antigenic stress that impinges throughout life upon innate immunity, and has potential implications for the onset of inflammatory diseases. Bulati points out some further important factors related to ageing:

- There is evidence of neuromuscular degenerative disease and other tissue dysfunction.
- Micronutrient inadequacy leads to metabolic consequences, e.g. DNA damage, cancer, severe infection, cognitive dysfunction and accelerated ageing.
- Reactivity of dendritic cells to self antigens can be characteristic of ageing. Furthermore, this over-reactivity induces T lymphocyte proliferation with subsequent higher risk of autoimmune diseases.
- Hyper activated T cells are possibly involved in bone loss associated with vascular disease in aged mice.
- There is a decrease in vaccine responsiveness leading to mortality.

De Martinis [7] and Franceschi [5] state that the most important characteristics of immunosenescence are the accumulation of memory and effector T cells, a reduction of naive T cells, shrinkage of the T cell repertoire and filling of immunological space. He points out that:

- The filling of the immunological space with memory and effector cells is a consequence of exposure to a variety of antigens over time.
- Clonal expansion of peripheral T cells carrying receptors for single epitopes of the herpes viruses Cytomegalovirus and Epstein-Barr virus are common in the elderly and are associated with a loss of early memory cells, an increase of T cytotoxic cells, a gradual filling of immunological space and an immune risk phenotype. The immune risk phenotype is a set of bioparame-ters associated with poor immune function. These parameters are: low levels of B cells, increased levels of cytotoxic T cells, poor T cell proliferative response, a T helper - T cytotoxic ratio of less than 1 and cytomegalovirus seropositivity [8].
- With the decline of immune function there is an increase in autoantibody frequency. An important result of this may be a loss of the ability to distinguish between self/nonself molecules.
- The lifelong respiratory burst, i.e. a reactive oxygen species causes damage to important cellular components (lipidic membranes, enzymatic and structural proteins and nucleic acids) during ageing. Oxidative damage is counteracted by several genetically controlled enzymatic and non-enzymatic antioxidant defence systems. All these protective mechanisms tend to become less effective with age.
- An elderly immune system becomes more predisposed to chronic inflammatory reactions and less able to respond to acute and massive challenges by new antigens. This continuous attrition is caused by clinical and sub-clinical infections, as well as the continuous exposure to other types of antigens, is likely responsible for the chronic innate immune system activation and inflammation.
- Inflamm-ageing, the peculiar chronic inflammatory status which characterizes ageing, is under genetic control and is detrimental to longevity. It leads to long term tissue damage and is related to an increased mortality risk.
- The unavoidable chronic overexposure to stress factors determines a highly pathogenic sustained activation of the stress-response system leading to a progressively reduced capacity to recover from stress-induced modifications.

## 3 Candidates for Immunosenescence Models

We decided to select the four most influential theories from the above as possible candidates for building computationally predictive systems and will now discuss them in more detail. The four theories are: a lack of naive T cells, immunological space filling with memory cells, innate system up-regulation and accumulation of T-regulatory cells.

### 3.1 Lack of Naive T Cells

Before an individual reaches the age of 20, the set of naive T cells is sustained primarily from thymic output [3]. However, in middle age there is a change in the source of naive T cells: as the thymus involutes, there is a considerable shrinkage in its T cell output, which means that new T cells are mostly produced by peripheral expansion. There is also a belief that some memory cells have their phenotype reverted back to the naive cells type [3].

However, these two new methods of naive T cell repertoire maintenance are not effective [3] as they do not produce new phenotypic changes in the T cells. Rather, evidence shows that they keep filling the naive T cell space with copies of existing cells. Therefore, the loss of clones of some antigen-specific T cells becomes irreversible. These age-related phenomena lead to a decay of performance in fighting aggressors.

### 3.2 Space Filling

The immune system deteriorates with age by losing functionality and immunocompetent cells. Moreover, it becomes limited in its use of resources. There is a finite number of T cells in operation at any time and to work properly, the immune system needs a reserve of naive T cells to combat new intrusions, and memory cells for previously encountered antigens.

With age, the repertoire of naive T cells shrinks proportionately to the previously faced threats, while memory cell numbers increase [3,7,5]. Late in life the T cell population becomes less diverse and some antigen-specific types of T cell clones can grow to a great percentage of the total T cell population, which takes up the space needed for other T cells, resulting in a less diverse and ineffective immune system. At some point there are not enough naive T cells left to mount any sort of effective defence and the total repertoire of T cells is filled with memory cells.

### 3.3 Innate Up-regulation

With age there is a decay in adequate functioning of the main phagocytes, i.e. macrophages, neutrophils [9] and dendritic cells [10]. As a consequence, deregulated immune and inflammatory responses occur in old people.

The investigation into the cellular and molecular mechanism underlying these disorders has provided compelling evidence that up-regulated cyclooxygenase and its product, particularly prostaglandin, play a critical role in the age-associated disregulation of the immune and inflammatory responses [6]. Increased prostaglandin production in old macrophages contributes to the suppression of T cell function with ageing. Furthermore, interventions targeted at decreasing prostaglandin production have been shown to enhance T cell-mediated function [6].

Thus, innate immunity and a high capacity to mount a strong inflammatory response, which is useful at younger age can become detrimental later in life. Inflamm-ageing can thus be considered a main phenomenon responsible for major age-related diseases and the evolutionary price to pay for an immune system fully capable of defending against infectious diseases earlier in life.

### 3.4 Accumulation of $T_{reg}$ Cells

The individual's ability to mount an effective immune response can be limited by regulatory elements such as significant changes in the number of T regulatory ($T_{reg}$) cells [11]. $T_{reg}$ cells act to suppress activation of the immune system and thereby maintain immune system homoeostasis and tolerance. The accumulation of $T_{reg}$ cells in old people inhibits or prevents some immune responses, e.g.

anti-tumoural ones. Also, the reduction of $_{Treg}$ cells might compromise the activation of immune responses in the aged. Therefore, an imbalance in $_{Treg}$ normal functioning can predispose immune dysfunction. This results in a higher risk of immune-mediated diseases, cancer or infections.

### 3.5 Discussion

A summary of the main characteristics of the candidate models described in the previous sections is presented in Table 1.

| | Theories | | | |
|---|---|---|---|---|
| **Characteristics** | Lack of Naive | Space Filling | Innate Up-regulation | $_{Treg}$ Cells |
| *Shrinkage of naive cells* | x | x | | |
| *Diversity decrease* | x | x | | |
| *Few clones taking space* | x | x | | |
| *Excessive memory cells* | | x | | |
| *Loss of clones* | x | | | |
| *Inflammation* | | | x | x |
| *Excessive T suppression* | | | x | x |
| *Degeneration* | x | x | x | x |
| *Auto-immunity* | x | x | x | x |
| *Less vaccine response* | x | x | | x |

**Table** 1. Main characteristics of the candidate models.

Recent research [2] states that the stimulation of high-avidity T cell receptors responses is essential for effective anti-tumour vaccines. High-avidity responses are capable of efficient anti-tumour activity in vitro and in vivo. However, in order to have effective vaccinations, there should be sufficient T cell response around tumours.

As we have discussed, when the organism ages it lacks T cells. Thus, we would like to know how critical immunosenescence is for T cell responses to anti-tumour vaccination in cancer treatments. In order to proceed with this study, we have chosen the first candidate model of immunosenescence, because it has the most direct focus on T cells which have a direct relation to cancer immunocompetence.

The number and phenotypical variety of naive T cells in an individual represents one of the main factors highlighted as influence in the process of immunosenescence. This number changes with age in quantity and diversity. It is also one of the first immune-components to show signs of ageing. Therefore, we believe that T cell responses to cancer might deteriorate with age. By understanding how T cell populations change over the years, we could get insights of what kind of T cell response there will be inside tumoural sites and towards vaccination.

In order to understand T cells dynamics with time, we developed an agent based simulation model presented in the next section. The simulation is based on data and equations obtained in [3]. The future objective is to use this agent-based model as an input for another model involving interactions of T cells and cancer.

## 4 Model

In this section we present the conceptual model together with the mathematical model and show how we transformed this into the agent based simulation that we performed.

### 4.1 Naive T Cell Output

A good indicator of thymic contribution to naive T cell output in an individual is the level of a particular biological marker called 'T cell receptors excision circle' (TREC). TREC is some circular DNA formed during the coding of T-cell receptors. The TREC percentage on a T cell decays with shrinkage of the thymic output and with the activation and reproduction of naive T cells [3]. This means that naive T cells originating from the thymus have a greater percentage of TREC than those originating through other proliferation.

Our model proposed here is based on data and equations obtained from [3], which is concerned with understanding naive T cell repertoire dynamics. The objective of Murray's model is to determine the likely contribution of each source of naive T cells, by comparing estimates of the presence of TREC in these cells (see Figure 1). The dynamics of the sustaining sources, i.e. naive proliferation, TREC and reversal of memory to naive T cells are each modelled mathematically.

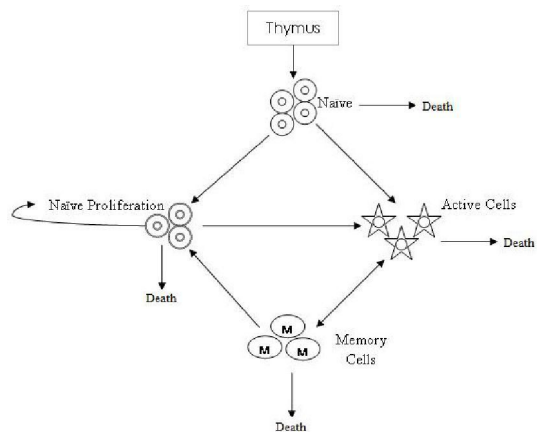

**Fig. 1.** Dynamics of Naive T cells.

## 4.2 The Mathematical Model

The mathematical model proposed in [3] is described by the equations (1) to (7) below. In these equations, $N$ is the total number of naive cells of direct thymic origin, $N_p$ is the number of naive cells that have undergone proliferation, $A$ is the number of activated cells, $M$ is the number of memory cells and $t$ is time (in years). The first differential equation is:

$$\frac{dN}{dt} = s0\, e^{-\lambda_t t} s(N_p) - [\lambda_n + /1_n g(N_p)]N, \quad (1)$$

where: $s0$ is the thymic output ; $\lambda t$ is the thymic decay rate, $s0\, e^{-\lambda_t t} s(N_p)$ represents the number of cells that arise from the thymus where $s(N_p)$ is the rate of export of the thymus defined by:

$$s(N_p) = \frac{1}{1 + \frac{\bar{s}\bar{N}_p}{N_p}} \quad (2)$$

As in the beginning of life there are no naive cells from proliferation, i.e. $N_p$, this parameter will have no influence on the output of the s function and therefore also no influence on the death rate of naive cells from the thymus. This can be better understood when we substitute $N_p$ by a number close to zero in the equation:

$$S(N_p) = 1/(1 + \bar{s}\bar{N}_p / \bar{N}_p) = 1/(1 + 0(number\_.0)/392) \approx 1.$$

Also, $\lambda_n N$ represents the naive cells' incorporation into the naive proliferating pool, where $\lambda_n$ is the naive proliferation rate, $/1_n$ is the thymic naive cells death rate, $/1_n g(N_p)N$ represents the naive cell death rate where the function $g(N_p)$ is the rate of change between naive TREC-positive and naive TREC-negative, defined as:

$$g(N_p) = 1 + \frac{\frac{bN_p}{\bar{N}_p}}{1 + \_\bar{N_p} \bar{N_p}}, \quad (3)$$

$\bar{N}_p$ and $\bar{s}$ are equilibrium and scaling values respectively.

For the simulation, s0 was defined as:

$$s0 = 0.82(7024 e^{(-((t-12\cdot 02)/3\cdot 623)^2)} + 5.203 \times 10^5 e^{(\_((t+127.8)/64.47)_2)} +$$

$$1937 e^{(-((t-7\cdot 357)/6\cdot 03)^2)} + 1.259 \times 10^{18} e^{(\_((t\_1309)/214.4)_2))}$$

The second differential equation is:

$$\frac{dN_p}{dt} = \lambda_n N + [ch(N, N_p) - /1_n]N_p + \lambda_{mn} M \qquad (4)$$

where: $c$ is the proliferation rate, $ch(N, N_p)N_p$ represents the naive proliferation where $h(N, N_p)$ is the dilution of thymic-naive cells through proliferation defined by:

$$h(N,N_p) = \frac{1}{1 + \frac{N+N_p}{N_p}} \quad (5)$$

$\mu_n N_p$ is the death rate of naive cells originated from proliferation and $\lambda_{mn}$ is the reversion rate from memory into $N_p$. The differential equation for memory cells (M) is:

$$\frac{dM}{dt} = \lambda_a A - \mu_m M - \lambda_{mn} M, \quad (6)$$

where: $\lambda_a$ is the reversion rate into memory and $\mu_m$ is the death rate of memory cells.

The final differential equation concerns the active cells (A) and is:

$$\frac{dA}{dt} = \lambda_{Na} N + \lambda_{Np} N_p - (\lambda_a + \mu_a)A. \quad (7)$$

where: $\lambda_{Na}$ is the activation rate of naive cells from the thymus (N). As we know, these cells have to proliferate before being activated. Therefore $\lambda_{Na}$ was set to zero. $\lambda_{NpA}$ is the activation rate of the naive cells from proliferation ($N_p$). $\mu_a$ is the death rate of active cells set to 44.4 given their short life span.

All rate values and other parameters for the model can be seen in Table 2. The values for these rates and parameters have been obtained by [3], through sensitivity analysis and parameter optimisation of the model. These parameter values, when used in conjunction with the above mathematical model, produced the results that most closely matched real-life observations of cell numbers[3]. Thus we use the same rates and parameter values in our model.

| parameter | value(s) |
|---|---|
| $\lambda_t$ | $\frac{log(2)}{15.7}$ $(year-1)$ |
| $\lambda_n$ | 0.003 |
| $\mu_n$ | 4.4 |
| $c$ | $\mu_n(1 - \frac{70}{N_p})$ |
| $\lambda_{mn}$ | 0 |
| $\mu_m$ | 0.05 |
| $\lambda_{Na}$ | 0 |
| $\lambda_{NpA}$ | 0.1 |
| $s$ | 0 |
| $\lambda_n$ | 0.003 |
| $\lambda_{mn}$ | 0 |
| $N_p$ | 392 |
| $b$ | 4.2 |

Table 2. Parameter values for the mathematical model.

In the next section, we will show how we can build an agent-based model representing equations (1) to (7) in order to investigate if it is possible to reproduce and validate the results obtained by [3].

### 4.3 The Agent Based Model

To convert the conceptual model into an agent based model, we first have to decide on what are the agents and their states: T cells are the agents and can assume four states, i.e. naive, naive from proliferation, active or memory. The agents' state changes and their death are defined by the ratios given in the mathematical model. Initially, all the agents are in the *naive* state. As the simulation proceeds, they can assume other stages according to the transition pathways defined in the conceptual model of figure 1.

The simulation scenario studied alters the function *g* over time by setting the parameter *b* greater than zero *(b* = 4.2). This means that the death rate of naive T cells from thymus will increase through the years as the number of naive cells from peripheral proliferation increases. There is no change of the thymic export, no reversion from memory to a naive phenotype and the conversion rate of naive cells from thymus to naive cells from proliferation is low (equal to 0.003). The simulation was run for a period of one hundred years and it considered the impact of thymic shrinkage per $mm^3$ of peripheral blood and 2000 initial naive cells from thymus.

The data collected for validation is plotted in the graph shown in Figure 2. The results from the simulation can be seen in Figure 3. The results show three different curves representing naive T cell populations derived from the thymus, peripheral naive T cells and the total number of naive cells. The curve representing naive cells from the thymus presents decay at the beginning of life followed by some interval of stability. By the age of twenty, the thymic export decreases following an exponential trend.

The results show that with the decay of naive cells derived from thymus, the naive repertoire changes from the thymic source to the peripheral proliferation source. And although the numbers of naive cells tend to be stable over time, there is no new phenotypical naive cell entering the system. This phenomenon reinforces the ideas of clone degeneration and space taken over by a few phenotypic clones.

As an extrapolation of the model, if we consider that around ten percent of the total T cell population is active in the organism during its lifetime, we can roughly get an idea of the behaviour of memory cells over time. This behaviour is shown in figure 4.

### 5 Conclusions

New trials to combat cancer are intending to take into account the individual patient's tumour biology and immune responses. *In vitro* stimulation of effector T cells has shown success in tumour treatments. To proceed with anti-tumour vaccinations *in vivo* it is necessary to determine the patient's immunocompetence profile. We believe that this profile and the patient's ability to fight cancer is influenced by immunosenescence and can be computationally modelled and simulated.

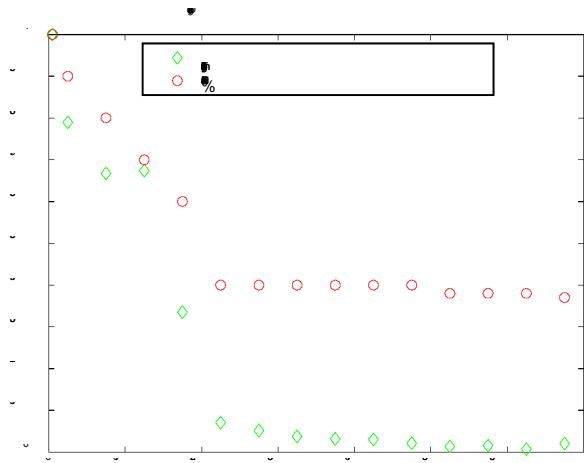

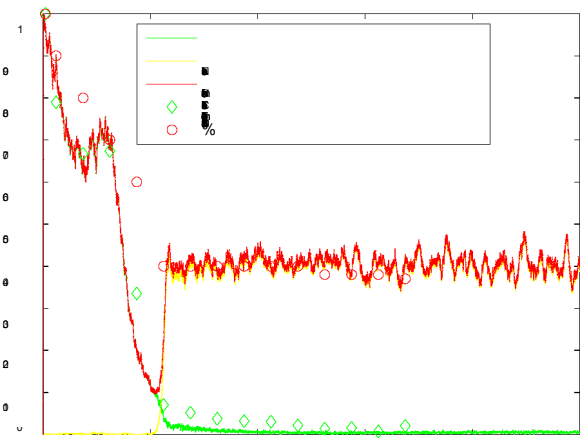

Fig. 2. The dataset used for validation was provided by [3] and [12].

Fig. 3. Results for the agent based simulation.

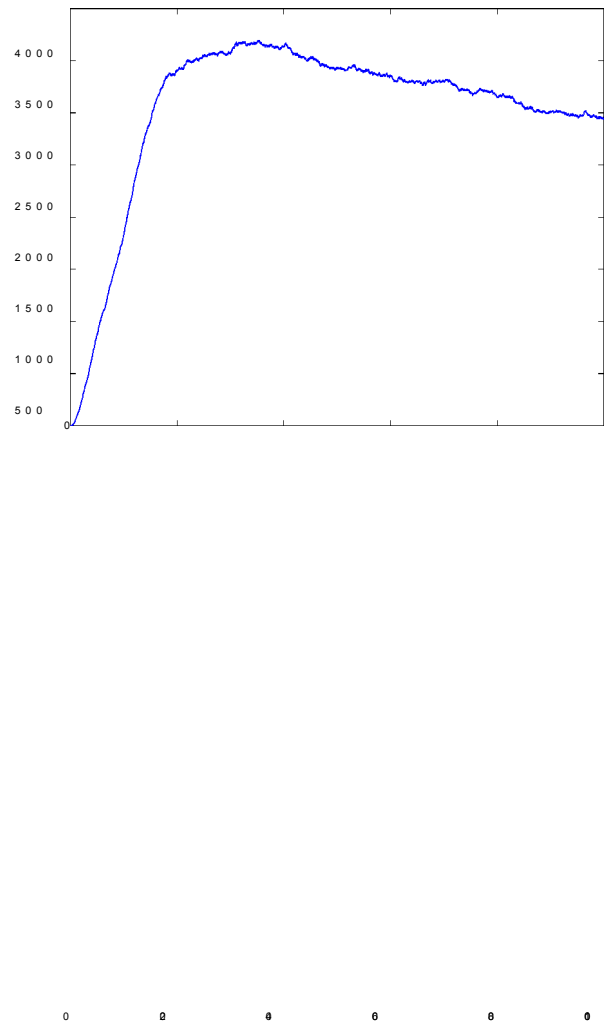

**Fig. 4.** Estimations of the amount of memory T cells with age.

We have shown in this paper how important factors related to immunosenes-cence that have a direct influence on T cell responses to cancer and vaccination can be investigated in a simulation model. In our research, the aim is to establish correlations between age, T cell populations and cancer immunocompetence. What we have done to date is to understand how T cell populations change with time.

The simulation model that we built and studied is based on mathematical equations converted into an agent based simulation. Results fit the observed data and thus we can estimate the likely contribution of each of the naive T cell repertoire maintenance method. With the decay of naive cells derived from the thymus, the naive repertoire changes from the thymic source to the peripheral proliferation source. The numbers of naive cells tend to be stable over time, but there is no new phenotypical naive cell entering the system. The lack of new T cell phenotypes probably makes vaccination and responses to cancer less effective. We believe that the results of this model could be used as an input to a bigger model, involving the movements of T cells inside tumours.

We could have built other types of simulation models based on the mathematical equations, and indeed have done so elsewhere, e.g. we have built a system dynamics model of the same problem for the 2010 Summer Simulation conference. Building a system dynamics simulation is a more straightforward task, as it readily lends itself to simulating differential equations. Agent-based models on the other hand require different conceptual models, including states and their transitions. This is non-trivial, but agent-based models do have the advantage of being more flexible. For instance, in future we intend to extend the model to include notions of space and movement (e.g. inside a tumour). Such additions are not easily possible when using other simulation paradigms. Moreover, agent-based simulations do not only work on an aggregate level, i.e. it is also possible to follow individual cells. We believe this point will hep us not only in our understanding and validation of future models, but also assist us in our explanations to non-computer scientists.

In order to move this research forward, the model needs to be extended to address the interactions between T cells and tumoural cells. We want to investigate how T cells would respond to vaccination and fight tumoural cells. We also want to look at the individual behaviour of T cells inside different types of tumours and cancers. Therefore, we believe that the choice of an agent-based model is the most suitable for our future work. Other future questions we intend to answer include the usability of the simulation by clinicians and which model would be more suitable for their understanding of the underlying immunology.

Specifically to continue the cancer model, we intend to use the actual mathematical model as part of a bigger set of equations representing T cell responses inside tumours. Data about T cell responses in colorectal cancer has been collected and will be used for a new mathematical model. Once we have established how T cells respond to cancer and stimulation, an additional factor will be added. We will consider the influence of age in quantity and quality of responses. We believe that this model could provide clinicians with information about the suitability and advantages of cancer vaccination in patients with a particular immunocompetence profile.